# Extreme anti-ohmic conductance enhancement in neutral diradical acene-like molecular junctions

Brent Lawson[1], Efrain Vidal[2,3], Michael M. Haley[2,3] & Maria Kamenetska[1,4,5]

[1]Department of Physics, Boston University, Boston, MA, USA
[2]Department of Chemistry & Biochemistry, University of Oregon, Eugene, OR, USA
[3]Materials Science Institute, University of Oregon, Eugene, OR, USA
[4]Department of Chemistry, Boston University, Boston, MA, USA
[5]Division of Material Science and Engineering, Boston University, Boston, MA, USA
e-mail: mkamenet@bu.edu; haley@uoregon.edu

**Abstract:**

We achieve, at room temperature, conductance enhancements over two orders of magnitude in single molecule circuits formed with polycyclic benzoquinoidal ($BQ_n$) diradicals upon increasing molecular length by ~5 Å. We find that this extreme and atypical anti-ohmic conductance enhancement at longer molecular lengths is due to the diradical character of the molecules, which can be described as a topologically non-trivial electronic state. We adapt the 1D-SSH model originally developed to examine electronic topological order in linear carbon chains to the polycyclic systems studied here and find that it captures the anti-ohmic trends in this molecular series. The mechanism of conductance enhancement with length is revealed to be constructive quantum interference (CQI) between the frontier orbitals with non-trivial topology, which is present in acene-like, but not in linear, molecular systems. Importantly, we predict computationally and measure experimentally that anti-ohmic trends can be engineered through synthetic adjustments of the diradical character of the acene-like molecules. Overall, we achieve an experimentally unprecedented anti-ohmic enhancement and mechanistic insight into electronic transport in a class of materials that we identify here as promising candidates for creating highly conductive and tunable nanoscale wires.

**Introduction**

With increasing drive for miniaturization, the properties of nanoscale materials such as individual particles or molecules take on added significance for the development of future technologies. For example, passing electronic current across single molecules attached to metal electrodes has been shown to result in switching or diode-like behavior, raising the prospect of molecules as active components in next generation electronic devices.[1–5] Appealingly, the properties of such molecular circuits can be tuned using synthetic control of the molecular atomic structure. A challenge in the field is that electronic states of an organic molecular bridge are typically off-resonance with the Fermi energy of the conducting electrons in the metal.[6–9] As a result, typical molecular junctions behave as quantum tunnel barriers and tend to be highly insulating, with an electronic conductance that decreases exponentially with molecule length. Thus, with few exceptions, molecules longer than ~2 nm in length have vanishing conductance and therefore possess limited usefulness as candidates for nanoscale electronics. Developing molecular materials that overcome this limitation, known as *ohmic* decay, is an important challenge for the field.

Recent work has identified radical molecules with open-shell electronic structure containing unpaired electrons, as molecular wires with topologically non-trivial electronic states that can sustain higher conductance over longer lengths.[10–15] At room temperature and ambient conditions, charged diradicals generated through chemical oxidation were shown to exhibit *anti-ohmic* behavior, with conductance increasing modestly with molecular length up to ~2 nm and then decreasing again.[11] This reversal to typical conductance decay in longer charged organic diradicals was attributed to decreased coupling of edge topological states across the extended molecular backbone due to their localized character and the rotational degrees of freedom of phenyl constituents in the molecules.[11,16] Incorporating a chain of charged radicals in series in a single molecule restored anti-ohmic trends at longer molecular lengths, but required challenging experimental protocols to maintain the high oxidation charged state of a single organic wire.[13]

Here, we report extreme anti-ohmic behavior at room temperature in neutral diradicaloid molecules composed of cyclic benzoquinoidal units linearly fused together as in acenes, which we term the $BQ_n$ series (Fig. 1). Critically, we observe that conductance is enhanced over two orders of magnitude for a molecular backbone length increase of ~5 Å from $BQ_1$ to $BQ_3$ even at low bias voltage. This is by far the highest enhancement observed to date and suggests at the unique potential of the neutral polycyclic diradicals to serve as efficient electron transport nanowires. These neutral systems are robust to oxidation and display increased anti-ohmic character at higher bias voltage. We measure the IV characteristics to reveal that the HOMO-LUMO energy gap of the $BQ_n$ series shrinks as expected with extended delocalization lengths in longer molecules and, critically, does not decay in the HOMO-LUMO gap as it does in most other 1D nanowires, such as the well-studied polyacetylene. We adapt and extend the 1D Su-Schrieffer-Heeger (1D-SSH) formalism to model the topological phases in cyclic planar aromatic compounds studied here and combine it with the non-equilibrium Green's function method to predict conductance trends. We demonstrate that the anti-ohmic behavior observed experimentally is intrinsic in this family of molecules when realistic bond-order parameters are incorporated into the 1D-SSH parametrization. Our analysis reveals that the conductance enhancement we observe is due to growing constructive quantum interference (CQI) between the frontier orbitals in fused cyclic systems. In contrast, in linear conjugated polymers like polyacetylene, destructive quantum interference (DQI) causes a drop in transmission probability in the HOMO-LUMO gap, resulting in near constant conductance with increasing molecular length as has been confirmed experimentally in recent studies. Finally, we leverage the tunability of the molecule system to demonstrate further enhancement of anti-ohmic behavior in a related set of benzothiophene-fused quinoidal molecules (BTQ$_n$, Fig. 1a) with higher diradical character than the BQ$_n$ series. Overall, our study identifies a new class of 1D molecular materials as superior nanowires and provides an analytical framework and mechanistic insights needed to guide the development of highly conducting molecular circuits.

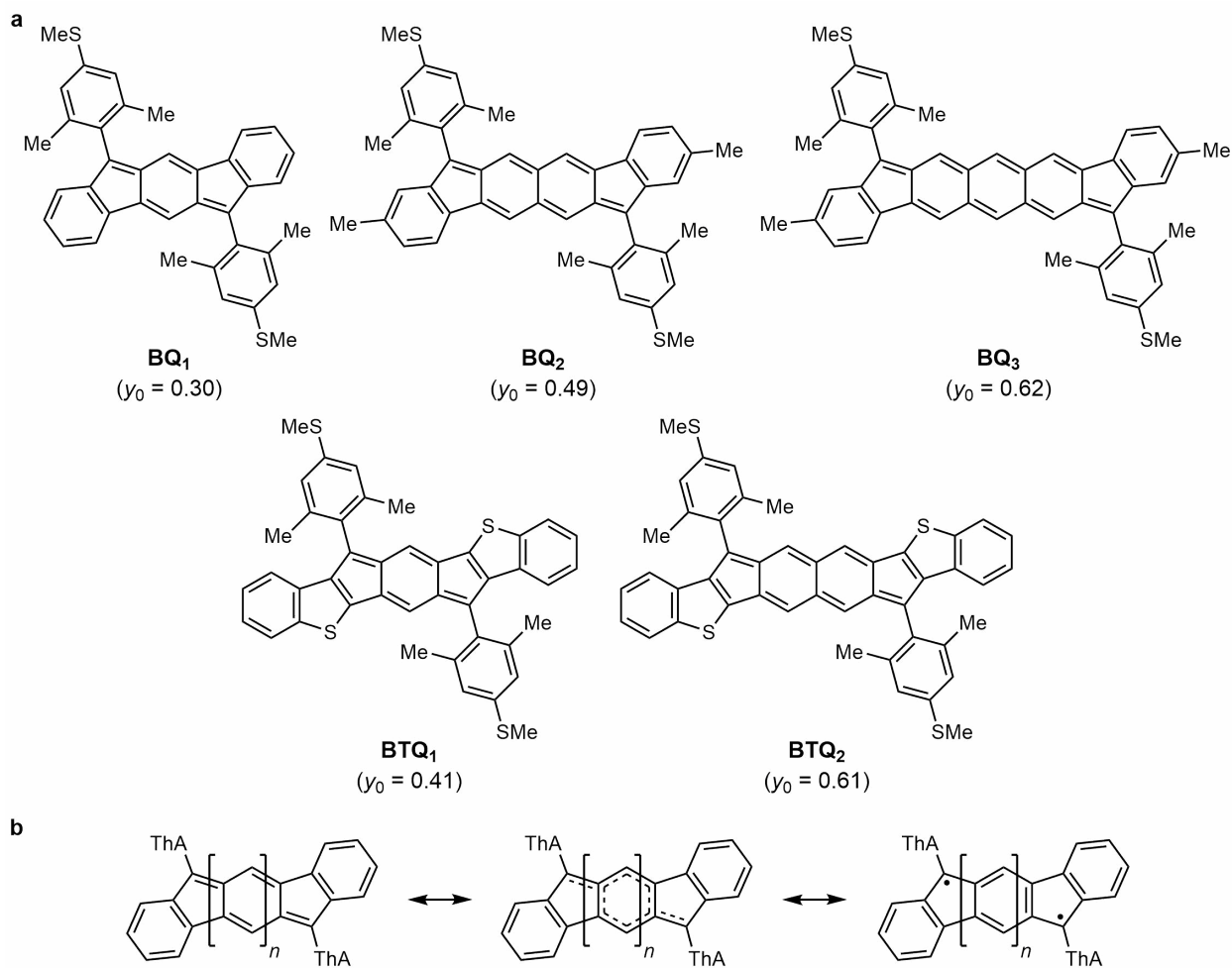

**Fig. 1| Molecular structures examined in this study and their resonance structures. a,** Structures of the **BQ** and **BTQ** molecules; calculated $y_0$ values (PUHF) are from Ref. 21. **b,** Resonance structures converting between the quinone, delocalized, and diradical form of the family of benzo-fused quinoids (**BQ$_n$**) where n=1-3. Thioanisole linkers (ThA = 2,6-dimethyl-4-(methylthio)phenyl) serve to anchor the molecule to gold electrodes in the junction.

**Results and Discussion**

We perform room temperature single molecule conductance measurements of the BQ$_n$ series using a scanning tunneling microscope break junction (STMBJ) technique.[17–19] The BQ$_n$ series shown in Fig. 1a are characterized by modest to intermediate diradical character, $y_0$, in their uncharged state.[20,21] Generally, as the number of fused six-membered rings increases, the diradical character of the molecule also increases, reflecting a higher residence of the molecules in the diradical structure (Fig. 1b, right) over the quinone (Fig. 1b, left)). Overall, these molecules have been shown to be air-stable open-shell diradical singlets in the ground state, with long term stability at room temperature.[20,22–26] Importantly, the diradical character $y_0$, which can be tuned rationally via judicious structural modification, is inherent to the molecule and independent of solvent conditions.[23]

The BQs and BTQs were prepared analogously to their previously reported mesityl analogues via the reaction of the lithiate prepared from 4-bromo-3,5-dimethylthioanisole (ThA) with the corresponding BQ and BTQ diones. Subsequent reductive de-aromatization with $SnCl_2$ afforded the target molecules; see the Supporting Information for the synthetic and spectroscopic details. Inclusion of the thioanisole functionality is critical to ensure binding to the Au electrodes. All measurements in the STMBJ were performed in 1 mM solution of 1-bromonaphthalene (BNP), a nonpolar organic solvent.

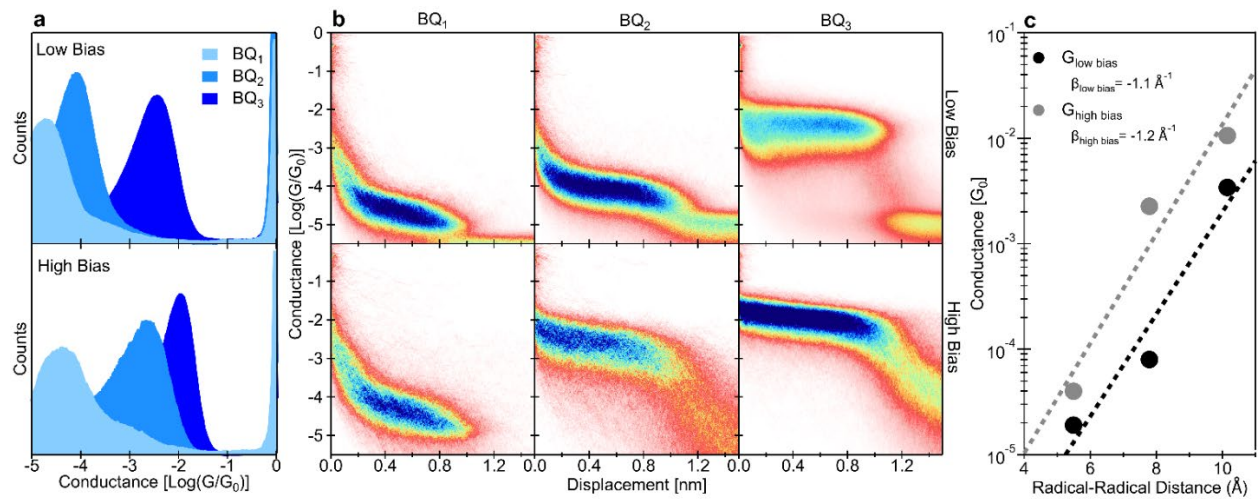

**Fig. 2| Conductance measurements of BQ$_n$ (n = 1-3). a,** Single-molecule conductance histograms of BQ$_{1-3}$ in BNP at low bias, –100 mV, in the top panel and high bias, –833 mV for BQ$_{1-2}$ and –250 mV for BQ$_3$, in the bottom panel. **b,** Single-molecule 2D conductance vs displacement histograms of BQ$_{1-3}$ at low bias in the top three panels and high bias in the bottom three panels. **c,** Calculated conductance from gaussian fits to the peaks in Fig. 2a for both low bias (black) and high bias (gray) with exponential fits given by the dashed lines.

The 1D and 2D conductance histograms constructed from more than 5000 traces recorded in the presence of molecules BQ$_{1-3}$ are shown in Fig. 2a,b. We compare conductance of the molecules measured at low bias of 100 mV, and high bias, > 250 mV, on the top and bottom of the figures, respectively. In both conditions, we observe distinct conductance peaks, attributable to the formation of stable single molecule junctions with each molecule. Importantly, we observe a conductance increase with increasing molecule length from 1 to 3 repeating quinone units. The most likely conductance of each molecule, determined from gaussian fits to the histograms, increases over two orders of magnitude upon a molecular backbone increase of ~5Å (Fig. 2c). Fitting these conductance trend to the expected Landauer model:[27]

$$G = G_c e^{-\beta L}$$

where $G_c$ is the contact conductance, $\beta$ is the conductance decay constant, and $L$ is the length of the molecules, we obtain a negative beta value of –1.1 Å$^{-1}$ for the low bias measurements. This measured value of $\beta$ < 0, indicates a reverse conductance decay and anti-ohmic transport trends in these molecules in the absence of chemical or electric oxidation.

Measurements at high bias, 833 mV for BQ$_{1-2}$ and 250 mV for BQ$_3$, show a further enhancement in single molecule junction conductance compared with measurements at 100 mV, especially for n = 2 and n = 3 (Fig. 2a, b bottom). The high bias voltages correspond to the conditions that can be sustained reliably by each molecular junction over several hours of data acquisition (see Supplementary Information for more details). We also observe that the distribution of measured conductance values for $BQ_3$ narrows at higher bias, which has been previously identified as a signature of near-resonant transport.[28] Increases in the conductance of $BQ_{2-3}$ at higher bias enhance the anti-ohmic characteristics, with a fit to the most likely conductance values yielding a $β$ of $–1.2$ $Å^{-1}$ (Fig. 2c); this corresponds to an unprecedented increase of nearly 3 orders of magnitude in conductance from n = 1 to n = 3.

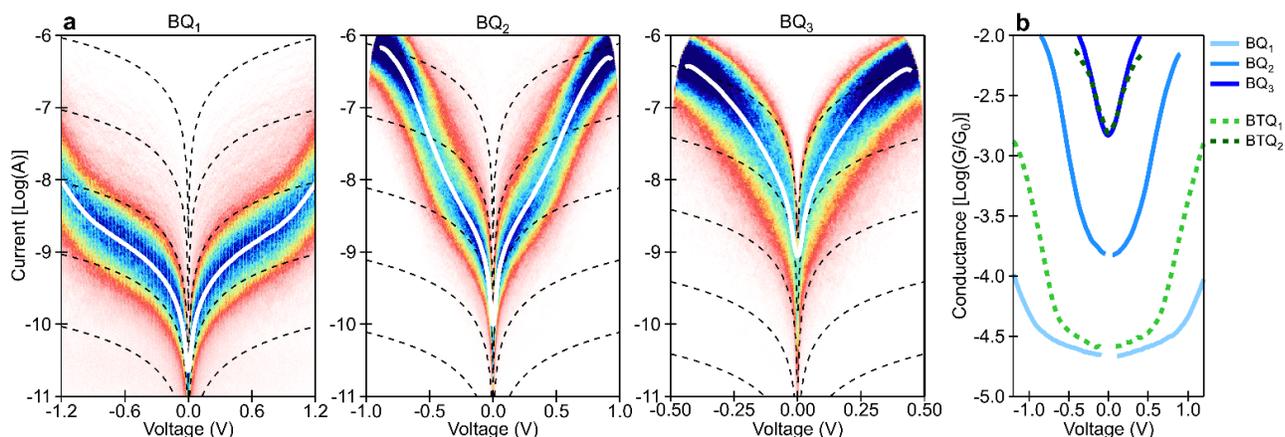

**Fig. 3| Current-voltage measurements of $BQ_n$. a**, Current-Voltage histograms for $BQ_{1-3}$ from thousands of IV curves taken while holding each molecule between two Au electrodes. Dotted lines correspond to the linear grid lines and the white lines are calculated from the average of gaussian fits of vertical slices of each histogram IV histogram. **b**, Fits of the conductance-voltage histograms derived from the current-voltage histograms for $BQ_{1-3}$ and $BTQ_{1-2}$.

To understand the voltage dependent conductance behavior of each molecule and estimate the position of the frontier electron orbitals relative to $E_F$, we perform single molecule current-voltage (IV) measurements by sweeping the voltage between the two electrodes while the molecule is held in the junction and recording the current using previously published protocols; additional details are provided in the Supplementary Information.[29] IV histograms for $BQ_{1-3}$ (Fig. 3a) are constructed from at least 5000 traces with the average IV curve, shown in white, calculated from gaussian fits to vertical slices at each voltage. The dashed black lines are linear curves shown for references. We observe that in all three cases, the molecular IV characteristics are non-linear at a range of biases probed here. For the shortest molecule, a linear current-voltage relationship is observed below ~500mV, but the current starts to increase exponentially above ~600 mV, suggesting that at this higher bias, the probed voltage range of ± 300 mV around $E_F$ approaches a resonance feature associated with a frontier electronic state in the transmission spectrum of $BQ_1$. In the longer molecules, $BQ_2$ and $BQ_3$, the exponential increase sets in above 100 and 0 mV, respectively, indicating the presence of molecular resonances within narrower voltage ranges for the longer molecules. Overall, the IV curves for the three molecules indicate a decreasing distance to one of the frontier orbitals as the number of repeating fused benzene units increases, suggesting a smaller HOMO-LUMO gap, $E_{gap}$, in longer molecules.

We replot the averaged IV curves from Fig. 3a as conductance and group them for direct comparison in Fig. 3b. The sharp upturns in these conductance spectra correspond to the supra-linear regions of the IV curves noted above. Importantly, we observe that the conductance at low bias, which probes transport in the vicinity of the $E_{gap}$ midpoint, follows anti-ohmic behavior. This reverse, anti-ohmic conductance decay at all bias regimes is anomalous compared to transport trends through most single molecule junctions observed to date. The data confirms that the neutral acene-like materials display anti-ohmic behavior in their neutral state, even in mild and accessible experimental conditions.

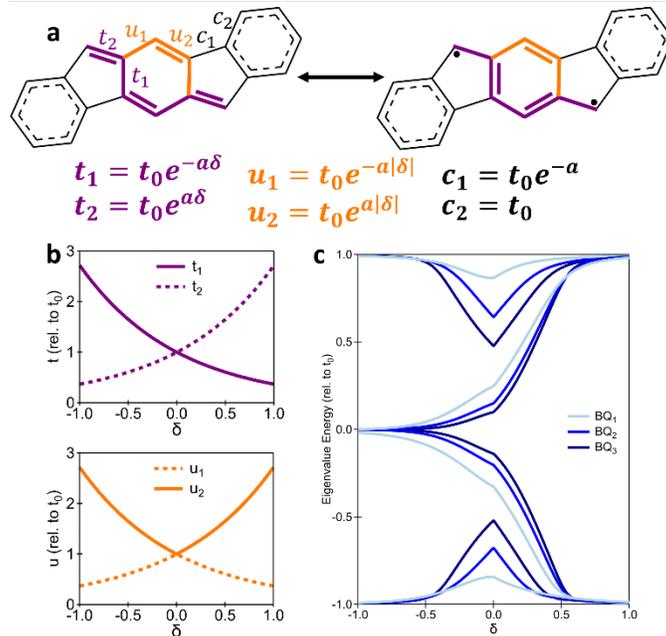

**Fig. 4| Modification to the 1D-SSH Model for polycyclic molecular systems like BQ$_n$. a**, Description of the coupling energies ($t_0$, $t_1$, $t_2$, $u_1$, $u_2$, $c_1$, $c_2$) describe by diradical parameter ($\delta$) and scaling parameter (a) used in the modified 1D SSH Hamiltonian. **b**, Graphical representation of the energies $t_1$, $t_2$, $u_1$, $u_2$. **c**, The calculated band structure for BQ$_{1-3}$ as a function of $\delta$.

We develop a 1D-SSH model appropriate for describing the contribution of the diradical character to the electron transport properties of quinonic molecular systems, such as BQ$_n$ studied here in order to probe the mechanism behind anti-ohmic transport. In prior studies, the 1D-SSH, or tight-binding, Hamiltonian was used to describe electronic phases in linear chains such as polyenes, which are molecules composed of a series of alternating single and double bonds, as shown in the purple path through BQ$_1$ (Fig. 4a).[16,30] In a Hamiltonian formalism, the double and single bonds correspond to a large and small coupling energy respectively.[7] For linear molecules, a single parameter $\delta$ was used to parameterize the transition between resonance structures, with $1 > \delta > -1$, spanning the range from zero diradical character $y_0 = 0$, to full diradical, $y_0 = 1$, respectively. For this parameterization, the coupling energy terms $t_1$ and $t_2$ are set to scale as $t_0 e^{\pm a\delta}$, with $t_0$ and $a$ set to 1, resulting in the purple bonds alternating between double (high coupling) and single (low coupling) as δ goes between –1 ($t_1 < t_2$ → diradical) and 1 ($t_1 > t_2$ → polyene).[7,16] Intermediate values of δ describe intermediate scenarios, with δ = 0 corresponding to a delocalized electronic state with 50% diradical character. As fully diradical polyenes (δ=-1) are not stable and thus experimentally inaccessible, this earlier model can instead be applied to cumulenes, which are linear carbon chains with a fully delocalized state well approximated by the δ = 0. In cumulenes and other molecular chains, conductance was found to remain nearly constant as the molecular length increased.[16,31]

Here, we implement a modified 1D-SSH Hamiltonian for polycyclic systems. We include an additional set of hopping parameters, $u_1$ and $u_2$, shown in orange (Fig. 4a), which are not entirely independent from $t_1$ and $t_2$. For example, unlike the purple bonds, the orange bonds do not flip from double to single in the transition from δ = –1 (diradical aromatic) to δ = 1 (closed-shell quinone). Instead, these bonds are described by alternative coupling terms ($u_1$, $u_2$) given by $t_0 e^{\pm a|\delta|}$ and they retain the same bond order value at δ = –1 as at δ = 1 (Fig. 4b). Critically, using this parameterization, all bonds $t_1$, $t_2$, $u_1$ and $u_2$ are equivalent when δ = 0 and the molecule is in a fully delocalized state. The delocalized benzene ring at the extrema of the molecules are described as constants ($c_1$, $c_2$) independent of δ, in agreement with previous experimental characterization of these systems.[20,23,25]

Using this model, we calculated the eigenvalues near $E_F$ as a function of δ (Fig 4c). As expected, the electronic structure transitions between the quinone (δ = 1) and diradical (δ = –1) states as the range of δ is scanned. As in the polyene system, here we observe a wide bandgap in the quinone structure, which closes as the molecule become a diradical.[32] These results confirm that our modification to the 1D-SSH model preserves the characteristic of the 1D topological insulators state at δ = –1, with $E_{gap}$ → 0 in this limit.

To model transport in the BQ$_n$ series, we convert the unit less 1D-SSH model above to a realistic energy scale characteristic of $sp^2$ systems containing alternating double bonds. We use established double-bond coupling energy parameters for graphene and the observed bond-orders in such systems to fit $t_0$ and $a$.[7,33,34] We obtain $t_0 = 2.7 eV$ and $a = 0.08$ (Supplemental Information). Establishing this pseudo-realistic 1D-SSH parameterization allows for the conversion of the reported diradical character $y_0$ to δ as $δ = -2y_0 + 1$. We note that the previously measured diradical character of BQ$_{1,2,3}$ correspond to δ = 0.4, 0 and –0.2, respectively (Fig. 5a).

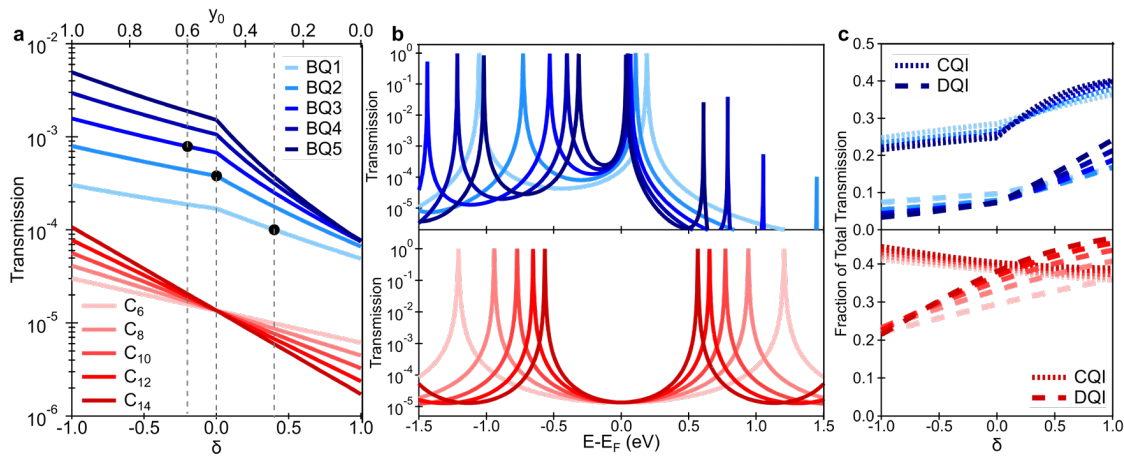

**Fig. 5| Anti-ohmic behavior of acene-like systems is unique and due to quantum interference effects. a**, Calculated transmission at E-E$_F$ = 0 eV of BQ$_n$ series and C$_n$ series as a function of diradical character. **b,** Transmission curves for BQ$_n$ series (top) and C$_n$ series (bottom) for the case δ = 0. **c**, Contribution of constructive quantum interference (CQI, dotted) and destructive quantum interference (DQI, dashed) to the total transmission for BQ$_n$ and C$_n$ series as a function of δ.

We use the wideband limit within the non-equilibrium Green's function (NEGF) approach described previously and in the Supplementary Information to calculate the transmission of molecules BQ$_n$ as a function of δ and compare them to the standard polyene (polyacetylene) case (Fig. 5a).[16,35] We calculate the polyene transmission using the standard 1D-SSH model with the realistic parameterization described above. In both cases, we observe that predicted conductance depends both on the diradical character δ and on the length of the molecular wire (Fig. 5a).

Importantly, however, we observe that in the case of the polycyclic quinoidal systems studied here and plotted in blue, transport is predicted to be anti-ohmic at *all* values of $\delta$. Our model predicts that even in the closed-shell $\delta = 1$ case, molecules of different length will follow reversed conductance decay. This is in direct contrast to the linear polyene chains, plotted in red, where transport is predicted to decay exponentially with molecule length for $\delta > 0$ and becomes anti-ohmic only in the experimentally inaccessible regime of $\delta < 0$. These predictions are consistent with experimental results measured in this manuscript and in prior work. For example, for polyenes, at $\delta = 0$, transmission is independent of length, which is consistent with experimental measurements of cumulene wires.[36,37] Similarly, for the BQ$_n$ series, characterized by an n and $\delta$ value marked by black points in Fig. 5a, the qualitative predictions of conductance enhancement match well to our experimental measurements. We also note that the essential features and trends in the transmission calculated using the 1D-SSH agree well with the more computationally expensive density functional theory (DFT)-based transmission calculations of relaxed junction geometries (Fig. S29), in agreement with previous theoretical calculations.[38] Our results suggest that the 1D-SSH one-electron model of the $\pi$ system captures the experimentally observed anomalous conductance enhancement in the BQ$_n$ series.[36]

Unlike many-body correlations, quantum interference (QI) among the frontier orbitals is included in the one-electron 1D-SSH formalism and is thus a candidate mechanism for the distinct conductance phenomena observed here in BQ$_n$ molecules. Previous work shows that QI effects can be explained by symmetry considerations of frontier MOs. To investigate the contribution of QI to transport trends in diradicals, we first compare the transmission spectra of polyenes and polycyclic systems at $\delta = 0$ calculated using the 1D-SSH (Fig. 5b). We observe that in both series, the E$_{gap}$ shrinks as the molecules grow longer, as expected from basic considerations of energy spacing with system size. However, for polyenes, the zero-bias conductance at $E_F$, remains constant, even as the band-gap closes, as was discussed above. This feature is a signature of destructive quantum interference (DQI) between the frontier orbitals.[9] In fact, ohmic conductance decay observed in the majority of polymeric molecular wires can be attributed to DQI, which counteracts the decrease in $E_{gap}$ at longer lengths to cause dips in transmission at $E_F$. In contrast, the predicted zero-bias transmission in $BQ_n$ grows with molecule length concomitantly with the closing of the band-gap, suggesting that DQI is a less significant factor in the polycyclic diradical system.

We calculate the CQI and DQI contributions explicitly using methods reported previously using the output of the 1D-SSH model developed here (Fig. 5c).[9] We observe very distinct QI trends in the linear polyene and polycyclic molecular systems. In polyenes, the transition from anti-ohmic to ohmic conductance coincides with an increase in DQI and a decrease in CQI as $\delta$ goes from –1 to 1. Specifically, DQI (CQI) increases (decreases) in magnitude from ~20% (~45%) to ~40% (~35%) of the total transmission. At $\delta=0$, CQI and DQI approximately cancel. In contrast, for the $BQ_n$ series, the CQI contribution dominates at all regimes and grows from ~20% to ~40% with increasing $\delta$, while the DQI fraction increases slightly but remains below ~20%. These trends corroborate and explain the origin of the extreme anti-ohmic behavior observed in $BQ_n$ compared to other polymers with diradical character. Critically, we see that in the case of polycyclic systems, increasing diradical character coincides with increasing CQI, providing mechanistic insight into engineering conducting molecular wires.

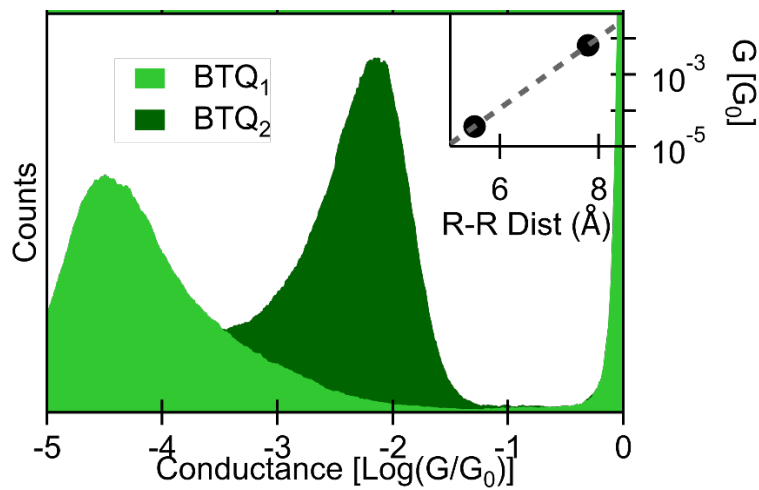

**Fig. 6| Engineering enhancement in anti-ohmic conductance through synthetic control.** Single-molecule conductance histograms of $BTQ_{1-2}$ (structures shown in Fig. 1a) in BNP at –500 mV. Inset: most likely conductance of $BTQ_{1-2}$ determined from the histograms plotted against radical-radical (R-R) distance results shows inverse decay and a β value of –2.3/Å.

To demonstrate the power of this experimental system, we measure the conductance of a derivative of the BQ series with benzothiophene elements fused to the benzoquinodal core with the thiophene S atoms in an *anti-* orientation to the diradical site (Fig 1a).[39] We term these the $BTQ_n$ and note that the backbone length of the quinoidal cores and structure of these molecules is identical to the $BQ_n$. $BTQ_n$ have been previously shown to have higher diradical character than $BQ_n$ for the same number of fused benzoquinone units n (Fig 1a). We measure the conductance of $BTQ_n$ for n = 1, 2 at a bias V = 500 mV (Fig. 6). We observe that conductance of this series is higher than in $BQ_n$ of identical length, particularly for the longer $BTQ_2$ molecule. The IV curves for the $BTQ_n$ series are plotted in Fig. 3b for direct comparison to $BQ_n$. Significantly, the conductance of $BTQ_2$ is enhanced compared to $BQ_2$ by at least 1 order of magnitude at all bias regimes, despite their identical radical-radical and overall lengths. We also note the inferred smaller $E_{gap}$ for BTQ$_2$ compared to BQ$_2$ as a result of increasing diradical character. Overall, we achieve an unprecedented reverse conductance decay in the BTQ$_n$ series, with a β value of –2.3 Å$^{-1}$ (Fig. 6 inset).

**Conclusion**

We have identified a unique class of molecular organic diradicals with a fused polycyclic core, where the equilibrium electronic structure is optimal for conductance enhancement at longer molecular lengths. We develop a 1D-SSH model which provides mechanistic insight into the source of conductance enhancement and identify quantum interference effects as key to anti-ohmic transport inherent to these systems. Crucially, we apply this understanding to demonstrate how molecular acene-like system can be synthetically altered and designed to further tune the diradical character, bandgap and QI to engineer molecular junctions with higher conductance and unprecedented anti-ohmic trends. These results set the stage for future engineering of conductance enhancement and other functional behaviors in cyclic diradicals based on an acene core.

**Methods**

STMBJ measurements are performed by repeatedly crashing and pulling out an Au tip from an Au substrate in the presence of molecules in solution, with a fixed bias across the two electrodes while recording the current, voltage, as well as the electrode displacement. Conductance is calculated at every displacement point during pull out by dividing current by voltage to generate a conductance trace. Plateaus in the conductance trace indicate formation of persistent junction configurations with a well-defined conductance. We statistically analyze the measurements by constructing 1D conductance histograms and 2D conductance vs displacement histograms from thousands of individually recorded raw data traces as detailed previously.[19,40,41] Reproducible molecular conductance plateaus result in peaks in the histograms in the conductance range below 1 $G_0$.

**Data availability**

All data supporting the findings of this study are available within the Article and its Supplementary Information, or from the corresponding author upon reasonable request.

**Acknowledgements**

MK and BL were supported by the US National Science Foundation (CHE-2145276 to MK). The synthetic part of this study (EF and MMH) was supported by the US National Science Foundation (CHE-2246964 to MMH).

**Author contributions**

M.K. and M.M.H. conceived the project. E.V. synthesized and characterized the samples. B.L. performed the single molecule measurements, developed the modified 1D-SSH model and carried out the calculations. B.L. and M.K. wrote the paper with key contributions from E.V. and M.M.H. The manuscript reflects the contributions and ideas of all authors.

**Competing interests**